\documentclass[showpacs,twocolumn,prb]{revtex4}
\usepackage{graphicx}
\usepackage{epsfig}
\usepackage{amssymb}

\def\beq{\begin{equation}}
\def\eeq{\end{equation}}
\def\beqa{\begin{eqnarray}}
\def\eeqa{\end{eqnarray}}

\def\e{\epsilon}

\def\D{\Delta}
\def\del{\delta}
\def\e{\varepsilon}
\def\cH{{\mathcal H}}

\def\hcon{\mathrm{H.c.}}

\def\L{\mathrm{L}}
\def\R{\mathrm{R}}

\def\al{\alpha}

\def\etal{{\sl et al.}}

\def\cdag{c^{\dagger}}

\def\nonum{\nonumber \\}

\def\del{\delta}

\def\nonum{ \nonumber \\}
\renewcommand{\sec}[1]{\vskip 0.truecm \noindent \emph{#1}. -- }

\begin{document}

\title{Thermo-spin effects in a quantum dot connected to ferromagnetic leads}
\author{Yonatan Dubi and Massimiliano Di Ventra}
\affiliation{Department of Physics, University of California - San Diego, La Jolla, California 92093-0319, USA} 
\begin{abstract}
We study a system composed of a quantum dot in contact with ferromagnetic leads, held at different temperatures. Spin analogues
to the thermopower and thermoelectric figures of merit are defined and studied as a function of junction parameters. It is shown
that in contrast to bulk ferromagnets, the spin thermopower coefficient in a junction can be as large as the Seebeck coefficient,
resulting in a large spin figure of merit. In addition, it is demonstrated that the junction can be tuned to supply only spin
current but no charge current. We also discuss experimental systems where our predictions can be verified.

\end{abstract} \maketitle
\sec{Introduction} Thermoelectricity - the relation between a
temperature bias and a voltage bias - is a very old problem of
solid-state physics. It has gained renewed interest in recent years
\cite{Mahan,Majumdar,book} due to the prospect of utilizing
nanostructures to develop high efficiency thermoelectric converters
\cite{Hochbaum,Boukai}. Theoretical models have been put forward for
the thermoelectric transport through quantum point contacts,
\cite{QPCtheory,ourwork} quantum
dots,\cite{QDtheory,Krawiec,Subroto}, molecular junctions
\cite{MolJuncTHeory} and other strongly correlated nanostructures
\cite{Freericks}.

Recently, Uchida \etal \cite{Uchida} have measured the spin equivalent of the charge Seebeck coefficient, namely the \emph{spin}
Seebeck effect, in which a temperature difference between the edges of a bulk ferromagnetic (FM) slab induces a spin-voltage
difference and generates spin current. These authors have suggested using this effect to construct a spin-current source for
spintronic devices \cite{spintronics}. However, the spin-Seebeck coefficient in this experiment was measured to be four orders of
magnitude smaller than the charge Seebeck coefficient. In addition, the temperature difference unavoidably generates also a
regular voltage bias across the sample, which may preclude easy applicability in spintronic devices.

Here we study the {\em thermo-spin effect}, i.e., the spin-analogue to the Seebeck effect in a nanojunction composed of a quantum
dot (either a molecule or a semi-conductor quantum dot structure) placed between two FM leads. The charge transport properties of
such systems have been studied both theoretically \cite{FQDFtheory} and experimentally \cite{FQDFexp1,FQDFexp2}. We define the
spin analogues of the thermo-electric coefficients, and show that in this particular case the spin- and charge-Seebeck
coefficients are of the same order of magnitude. We also calculate the thermo-spin figure-of-merit (FOM), and show it to be
relatively large, indicating high heat-to-spin-voltage conversion efficiency. Finally, it is demonstrated that the system
parameters can be tuned such that a large spin-current can be generated without any charge current, thus making this system ideal
for spintronic applications.

\sec{Definitions of spin-thermal coefficients} Consider a system
composed of some structure (for instance a quantum dot) placed
between two FM leads, which we assume have the same magnetization
alignment. The system is held at a temperature $T_\L=T_\R=T$. In
linear response, the thermo-electric Seebeck coefficient is defined
as minus the ratio between the voltage bias $\del V$ and the applied
temperature bias $\del T$ that generates it (in the absence of
charge current). In a spin system out of equilibrium one can define
a spin-voltage bias as $\del V_s=\D\mu_\R-\D \mu_\L$, where $\D
\mu_\nu=\mu_{\nu \uparrow}-\mu_{\nu \downarrow}$, with $\mu_{s\nu}$
the electro-chemical potential of the spin-$s$ on either right or
left of the the quantum dot. We expect this bias to be essentially
zero when measured in the bulk of the FMs, but it may acquire a
finite (albeit possibly small) value in proximity to the quantum
dot.

To derive the spin-Seebeck coefficient, we consider a system in
which there is both an infinitesimal temperature bias and
spin-voltage bias. The charge and spin currents are defined as
$I=I_\uparrow+I_\downarrow, ~I_s=I_\uparrow-I_\downarrow$,
respectively (note that
they have the same dimensions). In linear response, the spin-current is given by  $ I_s=G_s\del V_s+L_{T} \del T~~$%,\label{Is} $
where the response coefficient $L_T$ is related to the fact that a temperature gradient can induce both a spin flow and energy
flow~\cite{book,rem1}. Setting $I_s=0$, we find the spin-Seebeck coefficient \beq S_s=-\frac{\del V_s}{\del
T}=\frac{L_T}{G_s}~~.\label{Ss}\eeq Once $S_s$ and $G_s$ are defined, one may define a spin-FOM, $Z_sT=\left|\frac{G_s
S^2_s}{\kappa /T}\right|$, where $\kappa=\kappa_e+\kappa_{ph}$ is the thermal conductance of the system, which has an electron
contribution and a phonon contribution. The absolute value is taken because the spin-conductance $G_s$ may be negative. In
analogy with charge transport, one expects that a system with $Z_s T> 1$ is a good heat-to-spin-voltage converter\cite{Mahan}.
%\begin{figure}[t]
%\vskip 0.5truecm
%\includegraphics[width=6truecm]{fig1.eps}
%\caption{Schematic description of the model system, composed of a quantum dot between two FM leads. The leads are
%held in a spin-voltage bias $\del V_s$ and a temperature bias $\del T$.}\label{fig1}
%\end{figure}

\sec{Model} The model consists of a quantum dot between two FM leads. The corresponding Hamiltonian of the system is \beqa \cH
&=&\sum_{k,\nu=\L,\R} \sum_{s=\uparrow,\downarrow} (\e_{ks}-\mu_{\nu})\cdag_{\nu ks} c_{\nu ks}+ \sum_{s=\uparrow,\downarrow}\e_s
d^\dagger_s d_s  \nonum & &+U \hat{n_\uparrow} \hat{n_\downarrow}+ \sum_{k,\nu=\L,\R} \sum_{s=\uparrow,\downarrow}( \gamma_{\nu
ks} \cdag_{\nu ks}d_s+\hcon) ~~,\label{H} \eeqa where $\cdag_{\nu k s}$ creates an electron in the $\nu=\L,\R$ lead with spin $s$
and energy $\e_{ks}$ (the energy depends on spin due to the FM splitting) , $d^\dagger_s$ creates an electron in the dot with
spin $s$, $\hat{n}_s=d^\dagger_s d_s$ is the number operator, $U$ is the Coulomb charging energy, $\e_s$ is the energy level in
the dot, which is spin dependent due to a field-induced Zeeman splitting, $\Delta B$. The latter may originate from the
magnetic field induced by the FM leads or by an external field. It may also arise from the presence of spin-dipoles
 which are dynamically formed around a nanojunction \cite{Malshukov,book}. $\gamma_{\nu k s}$ is the coupling between the
leads and the dot. This is the simplest system that exemplifies the
physics of this paper but it can also be realized in experiments
\cite{FQDFexp1,FQDFexp2}.

 If the temperatures are higher than the Kondo temperature, in the sequential tunneling approximation (i.e. first order in
$\gamma_{\nu ks}$) one can describe the system by using rate equations \cite{Ralph}, which describe the populations of the
different states in the dot. The dot can be either empty (with probability $P_0$), populated by a spin-up electron ($P_1$), by a
spin-down electron ($P_2$) or by two electrons ($P_3$). The corresponding rate equation is \begin{widetext}\beq
\frac{d}{dt}\left(
\begin{array}{c} P_0 \\ P_1 \\ P_2 \\ P_3  \\ \end{array} \right) =\left(\begin{array}{cccc} -W_{0 \to 1}-W_{0 \to 2} & W_{1 \to 0} & W_{2
\to 0} & 0 \\ W_{0 \to 1}  & -W_{1\to 3}-W_{1 \to 0} & 0 & W_{3 \to 1} \\ W_{0 \to 2} & 0 & -W_{2 \to 3}-W_{2 \to 0} & W_{3 \to
2} \\ 0 & W_{1 \to 3} & W_{2 \to 3} & -W_{3 \to 1}-W_{3 \to 1} \end{array}\right) \left( \begin{array}{c} P_0 \\ P_1 \\ P_2 \\
P_3  \\ \end{array} \right)~~.\label{RateEquation}\eeq \end{widetext} The rates $W_{n\to n'}$ describe the probability per unit
time to transfer from state $n$ to state $n'$. They are evaluated by noting that the rate for an electron to hop onto (off) the
dot is proportional to the probability $p_{\nu s} (\e)$ to find an electron (hole) in the $\nu$-lead with spin $s$ at an energy
$\e$. We assume that the coupling between the leads and the dot is energy independent (wide band approximation) and for
simplicity assume that the leads are symmetric (it is easy to show that our results, e.g., Eqs.~(\ref{Ss}) and~(\ref{Ss2}) do not
depend on junction asymmetry). We thus have $p_{\nu s} (\e)=\al_s f_\nu(\e)$, where $f_\nu(\e)$ is the Fermi distribution of lead
$\nu$ (with the corresponding temperature and chemical potential). The constant $\al_s$ parameterizes both the dot-lead coupling
and the density of states (DOS), the spin dependence coming from both the FM band-shift and the tunneling rate\cite{Tedrow}. We
assume that there is a majority of spin up in the ferromagnets (and that the leads magnetizations are aligned), and define
$\gamma=\al_\downarrow /\al_\uparrow$. Thus, $\gamma$ encodes the difference between both the DOS and the tunneling rates of the
different spins.

Thus, for example, we have (setting $k_B=\hbar=1, \al_\uparrow=\al$ and $\al_\downarrow=\gamma \al$) \beqa W_{0 \to 1} &=& \alpha
[ f((\e_\uparrow-\mu_{\L})/T_\L)+f((\e_\uparrow-\mu_{\R})/T_\R)] \nonum
 W_{0 \to 2}&=& \alpha\gamma [ f((\e_\downarrow-\mu_{\L})/T_\L)+f((\e_\downarrow-\mu_{\R})/T_\R)] \nonum
 W_{1 \to 3}&=& \alpha \gamma [ f((\e_\downarrow+U-\mu_{\L})/T_\L)+f((\e_\downarrow+U-\mu_{\R})/T_\R)] \nonum
 W_{3 \to 2}&=& \alpha [(1-f((\e_\uparrow+U-\mu_{\L})/T_\L))+ \nonum & &~~+(1-f((\e_\uparrow+U-\mu_{\R})/T_\R))] ~~,
\label{rates} \eeqa and similarly for the rest of the transition rates. We assume that phonon-induced spin-relaxation processes
in the dot are inhibited due to the presence of FM leads, and are hence slower than the spin transfer time-scale and may be
neglected. We set the chemical potentials $\mu_L$ and $\mu_R$ as the zero of energy, and so the dot energies are
$\e_{\uparrow,\downarrow}=\e\mp 2 \mu_B\D B$ ($\mu_B$ is the Bohr magneton). We will discuss two limiting cases of small and
large Zeeman field $\D B$ (in the sense that $2 \mu_B\D B
>>\al$ or $2 \mu_B\D B<< \al$). The dot level $\e$ may be tuned, e.g., by a gate voltage.

The steady-state solution is obtained by equating the right-hand side of Eq.~\ref{RateEquation} to zero. From this solution, one
can determine the charge current, spin current and heat current, using the continuity equation. For the charge and spin currents,
one has $\frac{dn}{dt}=I_\R-I_\L$, and $\frac{dm}{dt}=I_{s\R}-I_{s\L}$ where $n$ is the charge on the dot and
$m=n_\uparrow-n_\downarrow$ is the magnetization of the dot. Using the rate equation one thus obtains (setting $e=\hbar=1$)
\begin{widetext} \beqa I_\nu&=&P_0 \left(W_{0\to 1}^{(\nu)}+W_{0\to 2}^{(\nu)}\right)-P_1 \left(W_{1\to 0}^{(\nu)}-W_{1\to
3}^{(\nu)}\right)-P_2 \left(W_{2\to 0}^{(\nu)}-W_{2\to 3}^{(\nu)}\right)-P_3 \left(W_{3\to
   1}^{(\nu)}+W_{3\to 2}^{(\nu)}\right) \nonum
I_{s\nu}&=&P_0 \left(W_{0\to 1}^{(\nu)}-W_{0\to 2}^{(\nu)}\right)-P_1 \left(W_{1\to 0}^{(\nu)}+W_{1\to 3}^{(\nu)}\right)-P_2
\left(W_{2\to 0}^{(\nu)}+W_{2\to 3}^{(\nu)}\right)-P_3 \left(W_{3\to
   1}^{(\nu)}-W_{3\to 2}^{(\nu)}\right)~,
    \eeqa\end{widetext} where $W_{n\to n'}^{(\nu)}$ are scattering rates of transitions between the dot and
   the $\nu=\L,\R$ lead. Once all the currents are obtained, it is a matter of algebra to obtain the different transport coefficients
   using the linear response definition of the spin current.

\sec{Results} The procedure described above allows us to obtain analytic expressions for all the currents and
thermo-electric/spin coefficients. The first result is that in the limit of $\D B \to 0$, the charge-Seebeck coefficient $S$ is
independent of $\gamma$ and is the same as was calculated in Ref.~\onlinecite{Subroto}. The spin-Seebeck coefficient $S_s$ is
found to be proportional to $S$,  \beq S_s=\frac{1-\gamma}{1+\gamma} S~~.\label{Ss}\eeq Thus, for normal leads ($\gamma=1$) we
have $S_s=0$, and for perfect FM leads ($\gamma=0$ or $\gamma=\infty$) the spin and charge coefficients are identical (up to a
sign). Eq.~(\ref{Ss}) shows that even for a moderate value of $\gamma=0.3$ we have $S_s \sim 0.5 S$, as opposed to the bulk case
where it is orders of magnitude smaller\cite{Uchida}. In the case of large $\D B$, the situation is even more interesting, since
in fact $S_s$ may become larger than $S$. In the limit of $U \to \infty$ and at $\e=0$ (i.e. the leads Fermi energies at the
center of the Zeeman splitting) we find that \beq \frac{S_s}{S}=\frac{2}{1-\gamma \exp \left( \frac{2 \mu_B \D B}{k_B T}
\right)}-1\label{Ss2}~~. \eeq For a value of $\D B=1.5$T at $T=5$K, a value of $\gamma=0.3$ yields $\frac{S_s}{S} \approx 2.6$.

\begin{figure}[h!]
\vskip 0.5truecm
\includegraphics[width=8truecm]{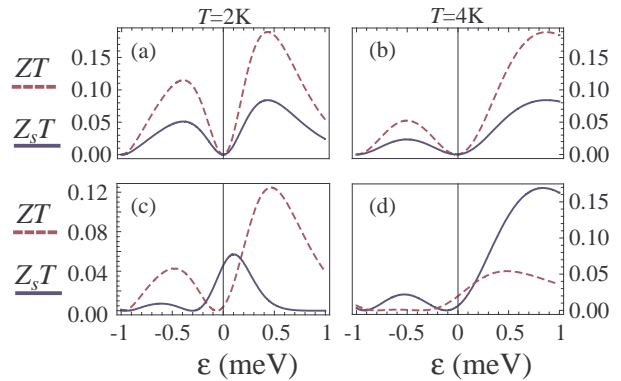}
\caption{(color online) Spin-FOM ($Z_sT$, solid line, blue online) and charge-FOM ($Z_T$, dashed line, purple online) as a
function of $\e$ at temperatures $T=2$K (left column) and $T=4$K (right column), for two values of Zeeman splitting, $\D
B=10^{-3}$T (a-b) and $\D B=1.5$T (c-d) (see text for other numerical values). The spin- and charge-FOM are comparable in size,
and at certain parameters spin efficiency may even exceed that of charge. }\label{fig1}
\end{figure}

Let us now turn our attention to the FOM. We have calculated the FOM (spin and charge) numerically. For this we take the
following parameters. The coupling between the dot and the leads is taken as $\al=10^{-2}$ meV (which is typical of semiconductor
quantum dots). The charging energy is taken to be two orders of magnitude larger, $U=2$ meV, and we take $\gamma=0.2$. We add to
the thermal conductance a phonon contribution which is $\kappa_{ph}=3 \kappa_0$ ($\kappa_0=\frac{\pi^2}{3}\frac{k_B^2}{h}T$ is
the quantum of thermal conductance \cite{quantum1,quantum2}), a reasonable value for nanoscale junctions \cite{Segal}. In
Fig.~\ref{fig1} we plot the spin-FOM ($Z_sT$, solid line, blue online) and charge-FOM ($ZT$, dashed line, purple online) as a
function of the dot energy level $\e$ for two temperatures of 2K and 4K (left and right columns, respectively) and for $\D
B=10^{-3}$T (Fig~\ref{fig1}(a-b)) and $\D B=1.5 $T (Fig~\ref{fig1}(c-d)). The first value is a typical field produced by regular
ferromagnets (e.g., iron), and the second corresponds to a large field splitting, which may be found in rare-earth ferromagnets
or be induced by an external magnetic field.

From Fig.~\ref{fig1} one  can see that the behavior of the spin and
charge FOM is similar for small Zeeman splitting, and that $Z_sT$
and $ZT$ are of the same order of magnitude. The situation is
different for large $\D B$, for which at certain energies close to
$\e=0$ one may obtain small $ZT$ but large $Z_sT$. This is due to
the fact that the Zeeman splitting in that case preserves the
particle-hole transport symmetry (the lack of which is responsible
for charge thermo-power) but dramatically changes the transport
properties of different spins, and hence increases the
spin-thermopower.

Finally, we study the system at finite currents. In the bulk, a
temperature gradient will inevitably induce both charge and spin
voltage \cite{Uchida}, and since the spin-Seebeck effect is much
smaller than the charge Seebeck effect, inducing large temperature
biases (to generate sizeable spin currents) would result in even
larger voltage biases. In the system studied here, one can instead
tune the system parameters such that there will be a large spin
current but vanishing charge current.

\begin{figure}[h!]
\vskip 0.5truecm
\includegraphics[width=6.5truecm]{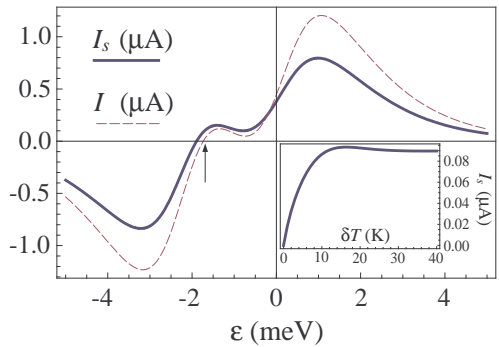}
\caption{(color online) Spin current (solid line, blue online) and charge current (dashed line, purple online) as a function of
$\e$ at a finite temperature difference $\del T=10 $K (see text for other numerical parameters). The arrow indicates the energy
at which the charge current vanishes and the spin-current is finite. Inset: spin-current as a function of temperature difference,
evaluated at the energy $\e$ at which the charge-current vanishes. }\label{fig2}
\end{figure}

In Fig.~\ref{fig2} the spin current $I_s$ (solid line) and charge
current $I$ (dashed line) are plotted as a function of $\e$. Here
the temperature $T=5$K, $\Delta B=1.5$T and we have added a constant
temperature gradient $\del T=10 $K (the spin- and charge- voltage
biases are zero). When the charge current vanishes (indicated by an
arrow in Fig.~\ref{fig2}) the spin current remains finite. The inset
of Fig.~\ref{fig2} shows the dependence of the spin-current,
evaluated by varying the energy $\e$ so that the charge current
vanishes, as a function of the temperature bias $\del T$. The
magnitude of the spin current increases with the temperature
difference, and attains significant values for realizable
temperature differences, until it saturates at large temperatures
(note, however, that the saturation temperature is comparable to the
interaction energy $U$, and hence one expects that the sequential
tunneling approximation breaks down at these temperatures). The
finite spin current at large temperature difference stems from the
fact that while the right lead is held at a high temperature, the
temperature in the left lead is still low, allowing for differences
in the tunneling rates of the different spins to be substantial. We
stress that a situation of finite $I_s$ but vanishing $I$ can not be
achieved by using only a voltage bias, but a temperature bias is
needed.

%\sec{Summary} In this communication we have evaluated the spin-thermopower and FOM for a system composed of a quantum dot coupled
%to FM leads, and showed that it may be as large (and even larger) as its charge counterpart, as opposed to the bulk case (where
%the spin-thermopower is four orders of magnitude smaller than the charge thermopower \cite{Uchida}). In addition, we have shown
%that one can tune the parameters to generate spin currents but no charge currents, a situation which can be achieved in the
%presence of a temperature difference.

Our results are valid even if one considers additional single
particle levels in the dot. In the limit of infinite $U$, in fact,
Eq.~(\ref{Ss}) is exact for the case of equidistant levels with no
Zeeman splitting. In the case with Zeeman splitting, we have
numerically estimated $S_s/S$ for up to five levels and found that
even in the presence of the additional levels $S_s/S\sim 1$.

The ability to couple a quantum dot to FM leads \cite{FQDFexp1,FQDFexp2} and to measure a local spin bias \cite{Uchida} have been
demonstrated experimentally. It is thus reasonable that the results presented here are accessible by future experiments. Another
interesting candidate for such experiments is graphene, for which both the possibility to fabricate quantum dots \cite{Graphene1}
and to bond FM leads to measure spin currents \cite{Graphene2} have been demonstrated.

We also point out that if the leads are FM, extracting the spin-current (or measuring the spin-voltage) has to be done close to
the junction, at a distance shorter than the spin-diffusion length of the FM leads. Possible ways to circumvent this difficulty
include the use of half-metallic leads (in which the spin-diffusion length should be very large) or to use a normal metal in
contact with a thin FM layer for each lead, with the FM thin layers sandwiching the quantum dot.
%
%It would be interesting to study the effects of correlations at low temperatures. This requires using a more elaborate
%calculation (using, e.g., non-equilibrium  Green's functions) and is beyond the scope of the present communication. It will also
%be of interest to study the effect of phonons and leads with opposite spin polarization on the spin-Seebeck effect. Such studies
%are currently underway.

We thank M. Krems and A. Sharoni for fruitful discussions. We are grateful to Yu. V. Pershin for crucial comments and to L. J.
Sham for valuable remarks. This work has been funded by the DOE grant DE-FG02-05ER46204 and UC Labs.


\begin{thebibliography}{}
\bibitem{rem1} There should also be a term proportional to $\del V$ (the charge bias) which is disregarded, since it does not supply any information
on the thermal efficiency.
\end{thebibliography}
\end{document}